\documentclass[11pt]{article} 
\usepackage{graphicx}
\usepackage{amsmath,multirow}
\usepackage{enumerate,array}
\usepackage{amsfonts}
\usepackage{amssymb}
\usepackage{epsfig}
\usepackage{graphics}
\usepackage{multicol, multirow}
\usepackage[pdfstartview=XYZ, bookmarks=true, colorlinks=true, linkcolor=blue, urlcolor=blue, citecolor=blue, pdftex, bookmarks=true, linktocpage=true, hyperindex=true]{hyperref}
\usepackage{enumitem}
\usepackage{color}
\usepackage{amsthm}
\usepackage{float}
\usepackage{lscape}
\usepackage{verbatim}
\usepackage{amsfonts}
\usepackage{lineno}

\newcommand{\beq}{\begin{equation}}
\newcommand{\eeq}{\end{equation}}

\makeatletter
\renewcommand\@biblabel[1]{}
\makeatother
\renewcommand{\baselinestretch}{1.2}

\begin{document}
	
	\title{Building nonstationary extreme value model using L-moments}
	
	\author{Yire Shin$^{1}$, Yonggwan Shin$^{2, *}$, %Sanghoo Yoon$^{1, 3}$,
			 Jeong-Soo Park$^{1}$ \\ 
			\small \it 1: Department of Statistics, Chonnam National University, Gwangju 61186, Korea\\
			\small \it 2: R\&D center, XRAI Inc., Gwangju 61186, Korea \\
		%	\small \it 3: Department of Statistics, Daegu University, Daegu 68453, Korea \\
%			\small\it 3: DIRC for Integrated Disaster Management in the Watershed, \\ 		
%			\small\it Mahasarakham University, Maha Sarakham 44150, Thailand. \\
			\small \it *: Corresponding author, email: syg.stat@gmail.com
		}
	\maketitle %\linenumbers
	
	\begin{abstract}
%%%%%%%%%%%%%% abstract %%%%%%%%%%%%%%%%%%%%%%%%%%%
The maximum likelihood estimation for a time-dependent nonstationary (NS) extreme value model is often too sensitive to influential observations, such as large values toward the end of a sample. Thus, alternative methods using L-moments have been developed in NS models to address this problem while retaining the advantages of the stationary L-moment method. However, one method using L-moments displays inferior performance compared to stationary estimation when  the data exhibit a positive trend in variance. To address this problem, we propose a new algorithm for efficiently estimating the NS parameters. The proposed method combines L-moments and robust regression, using standardized residuals. A simulation study demonstrates that the proposed method overcomes the mentioned problem. The comparison is conducted using conventional and redefined return level estimates. An application to peak streamflow data in Trehafod in the UK illustrates the usefulness of the proposed method. Additionally, we extend the proposed method to a NS extreme value model in which physical covariates are employed as predictors. Furthermore, we consider a model selection criterion based on the cross-validated generalized L-moment distance as an alternative to the likelihood-based criteria.
\end{abstract}
	
\vspace{5mm} \noindent {\bf Keywords}: {Cross-validated L-moment distance, Expected number of events, Solving nonlinear equations, Transformation to stationarity, Standard Gumbel distribution.}

%%%%%%%%%%%%%%% Introduction %%%%%%%%%%%%%%%%%%%%%%%%%%%
\section{Introduction}
	
The stationarity of time-series data is defined as a distribution that remains the same for any subsample of the original sample. However, natural phenomena rarely satisfy this hypothesis (Parey and Hoang 2021). In such cases, when a phenomenon evolves over time, a model-fitting technique that considers nonstationarity may be required. %This study focuses on developing an algorithm to estimate the nonstationary (NS) parameters efficiently in an extreme value model. 
	
Under stationarity, L-moment estimation (LME) often offers advantages over maximum likelihood estimation (MLE), including easy computation, robustness, and less bias for small and moderate sample sizes (Hosking and Wallis 1997; Jeong et al.~2014; Naghettini 2017). When estimating the shape parameter ($\xi$) and high quantiles of the generalized extreme value (GEV) distribution, LME outperforms MLE in terms of a lower root mean squared error (RMSE), especially when the upper endpoint is unbounded (Hosking et al.~1985; Yilmaz et al.~2021; Hossain et al.~2021). 

Applying LME is not as straightforward in the case of nonstationarity or multivariate situations, whereas applying MLE is straightforward in complex situations. Therefore, MLE has been predominantly employed to estimate parameters of nonstationary (NS) extreme value models, which have recently received more attention in the context of climatic change (e.g., Konzen et al.~2021; Bousquet and Bernardara 2021; Choi 2021; Baldan et al.~2022; Prahadchai et al.~2023; Radfar et al.~2023). However, MLE  is significantly influenced by outliers or large values toward the end of a sample, resulting in inferior performance in such cases. 

% Cannon (2010) developed a framework for the NS GEV model using a probabilistic extension of the multilayer perceptron neural network in conjunction with the generalized MLE method. Nasari et al.~(2017) and Rohmer et al.~(2021) presented spline models capable of capturing the behavior of NS nonlinear quantiles. %However, Nasari et al.~(2017) sometimes failed to demonstrate improved performance in extrapolating quantiles for high return periods. 
%In addition, Hoang (2010) considered a mean-variance trend approach by estimating the moment functions nonparametrically, serving as an appropriate alternative to parametric methods. The mentioned models are flexible and represent a wide range of NS relationships well. However, at least a moderate sample size may be required to perform effectively because the models rely on MLE or nonparametric smoothing. 
 
	Several studies have considered LME as a valuable and robust alternative to MLE to analyze time-dependent NS extreme values. Strupczewski and Kaczmarek (2001) considered the weighted least squares (WLS) method in which the parameters were estimated using the LME method for standardized residuals. However, the WLS estimator is sometimes too sensitively influenced by large values toward the end of a sample.
	Moreover, Ribereau et al.~(2008) and Mudersbach and Jensen (2010) considered L-moment-based methods to analyze time-dependent NS extreme values, which are computationally easier but perform well only under restricted assumptions.
	%  proposed a regression-based algorithm to encompass temporal dependencies using generalized probability-weighted moments. Compared to the MLE, this method is computationally easier and performs better but is only applied to cases where nonstationarity is expressed in terms of the location parameter. Further, Mudersbach and Jensen (2010) calculated the usual LME in a time-sliding window in which the data are considered stationary. These LMEs, depending on time, were fitted to an NS function using the nonparametric regression method. %Then, the NS GEV distribution (GEVD) is used to extrapolate high quantiles. This method is easy to implement but requires a local stationarity assumption within the time window. 

	Gado and Nguyen (2016a) proposed a method (called GN16 hereafter) based on the framework by Cunderlik and Burn (2003). %Time-dependent regression models of location and scale parameters were employed to detrend the time series of the annual maximum flood by Cunderlik and Burn (2003) and GN16 to ensure stationarity after detrending (see Section~\ref{GN16} for details). 
	The GN16 method, which obtained stationary L-moment estimates for the detrended sequences, performs well in many cases for the NS GEV model, except for positive trends in the standard deviation. %The GN16 method  obtained stationary L-moment estimates for the detrended sequences. %Then, they estimated the NS parameters using the relationships between the NS GEV parameters and temporal moments. 
	%The GN16 method also performs well in many cases for the NS GEV model, except for positive trends in the standard deviation.
	
This study focuses on developing a new algorithm to estimate NS parameters efficiently to overcome the weaknesses of the GN16 and WLS methods. The proposed method employs robust regression, transformation to stationary data, and LME. We expect the advantages of LME for a stationary model to continue to hold when estimating the parameters of NS models using the proposed method. %Moreover, we provide an extension of the proposed algorithm to NS GEV models with physical covariates employed as predictors.
	
Section~2 describes extreme value models and preliminaries, and Section~3 presents the existing L-moment-based methods for NS GEV models and the proposed method. Next, Sections~4 and 5 provide a simulation study and a real data application. %Section~5 illustrates the effectiveness of the proposed method by applying it to extreme streamflow data from Trehafod, UK. 
Section~6 extends the proposed method to the NS model with physical covariates. Finally, discussion and conclusion are provided in Sections~7 and 8. The accompanying Supplementary Information includes additional tables and figures. %The first version of the R code based on this proposal is available on GitHub at https://github.com/yire-shin/non\_stationary-gev.git.
	
\section{Extreme value models and estimation}
	
\subsection{Generalized extreme value distribution}
The GEV distribution (GEVD) has been widely used to model the extremes of natural phenomena and human society. The primary reason for applying GEVD is that it is supported by the large sample theory (Coles 2001). %\cite{coles2001introduction}. 
The cumulative distribution function (CDF) of the stationary GEVD is as follows (Hosking and Wallis 1997):
%\begin{linenomath*}\begin{equation} \label{pdf-gevd}
%	f(x) = \sigma^{-1} \left(1 -\xi {{x-\mu} \over {\sigma}}\right) ^{{1 \over \xi} -1}\times F(x), \end{equation}\end{linenomath*} where
\begin{linenomath*}\begin{equation} \label{cdf-gevd}
	F(x) = \text{exp} \left\{ - \left(1 -\xi {{x-\mu} \over {\sigma}}\right) ^{1/ \xi} \right\} ,
\end{equation}\end{linenomath*}
when $1-\xi (x- \mu ) / \sigma > 0$ and $\sigma>0$, where $\mu,\; \sigma$, and $\xi$ are the location, scale, and shape parameters, respectively. The case of $\xi=0$ in (\ref{cdf-gevd}) corresponds to the Gumbel distribution. It is important to note that the sign of $\xi$ in (\ref{cdf-gevd}) was changed from that in the work by Coles (2001) to follow the work by Hosking and Wallis (1997). 
{The reason why the sign of $\xi$ is changed in this study is that considerably many reports and softwares (Asquith 2011) based on L-moments, especially in hydrology, have employed the changed sign of $\xi$.
	Consquently, we remark again that the GEV distribution has different tails according to the sign of the shape parameter $\xi$; heavy tail when $\xi<0$, exponential when $\xi \rightarrow 0$, and light or bounded when $\xi>0$, where an opposite sign to Coles (2001) is used.}

We use the following notation: $Z_t \sim GEV(\mu_t,\; \sigma_t,\; \xi_t)$, for the time-dependent NS GEV model. For example,
\begin{eqnarray} %\label{mu_sigma_t}
\mu_t &=& \mu_0 + \mu_1 \times t , \label{mu_t} \\
\sigma_t& =& \text{exp}(\sigma_0 + \sigma_1 \times t), \label{sigma_t}
\end{eqnarray}
where we set $t = \text{year} - t_0 +1$, so that $t = 1,2,\cdots,n$, where $t_0$ is a starting year of the observations and $n$ is the sample size. An exponential function in (\ref{sigma_t}) is used to ensure that the positivity of $\sigma$ is respected for all $t$. The shape parameters ($\xi_t$) are challenging to estimate precisely; modeling $\xi$ as a function of time is typically unrealistic (Coles 2001, pp.~106; Katz 2013). The GEV model with the location and scale parameters represented by (\ref{mu_t}) and (\ref{sigma_t})  under constant shape parameter is denoted as the GEV11 model. 

Simpler models can be built with a constant scale parameter: $\sigma_t =\sigma_0$. The GEV model with the time-dependent location parameter (\ref{mu_t}) and constant scale parameter $\sigma_0$ is denoted as the GEV10 model in this study. When $\mu_t = \mu_0 + \mu_1 \times t + \mu_2 \times t^2 $, under constant scale and shape parameters, the model is called the GEV20 model. {Many other NS GEV models including the GEV11 and GEV20 models are considered in Coles (2001), Reiss and Thomas (2007), AghaKouchak et al.~(2013), Katz (2013), and Prahadchai et al.~(2023), for instance.} 

%The number of parameters in the GEV11 model is five, whereas the number of stationary GEVD parameters is three. Thus, the GEV11 model is more flexible and complex than the stationary GEVD. In general, the GEV11 model results in less bias but more variance in estimation than the stationary GEVD due to the bias-variance trade-off (e.g., James et al.~2013). This general rule is also observed in the Monte Carlo simulation study presented in Section~\ref{sim_study}.

\subsection{L-moment estimation}
Hosking (1990) introduced L-moments as a linear combination of the expectations of order statistics. The $r$th L-moments ($\lambda_r$) of a probability distribution are defined as follows: 
%by Hosking and Wallis (1997)
\begin{eqnarray} \label{l-moments}
\lambda_1 &=& E (X_{1:1} ), ~~~~
\lambda_2 = {1\over2} E (X_{2:2}- X_{1:2}), ~~~~
\lambda_3 = {1\over3} E(X_{3:3}-2 X_{2:3}+ X_{1:3}), \\
%\lambda_4& =& {1\over4} E(X_{4:4}-3 X_{3:4}+ 3X_{2:4} – X_{1:4} ),
\lambda_r &=& r^{-1} \sum_{j=0}^{r-1} \;(-1)^j \;{{r-1}\choose{j}}  \;E(X_{r-j:r}), ~~r > 3,
\end{eqnarray}
%and in general
%\beq
%\lambda_r = r^{-1} \sum_{j=0}^{r-1} \;(-1)^j \;{{r-1}\choose{j}}  \;E(X_{r-j:r}),  
%\eeq
where $X_{k:n}$ is the $k$th smallest observation from a sample of size $n$. The L-moment ratios are defined as follows: $\tau_2 = \lambda_2 / \lambda_1, ~ \tau_r = \lambda_r / \lambda_2, ~ r=3,4,...$
%\beq \label{L-ratios}
%\tau_2 = \lambda_2 / \lambda_1, ~~~ \tau_r = \lambda_r / \lambda_2, ~~ r=3,4,...
%\eeq
 The sample L-moments, denoted by $l_r$, were obtained as unbiased estimates of $\lambda_r$. Usually, $l_r$ is a linear combination of the ordered sample values $x_{1:n},\;\cdots, x_{n:n}$. The sample L-moment ratios are obtained as follows: $t_2 = l_2 / l_1, ~ t_r = l_r / l_2, ~ r=3,4,...$
%\beq
%t_2 = l_2 / l_1, ~~ t_r = l_r / l_2, ~~ r=3,4,...
%\eeq
%The L-moments offer theoretical and practical advantages over ordinary moments. For example, L-moments are less sensitive to outlying data and better identify the parent distribution that generates a particular data sample (Hosking and Wallis 1997).

Analogous to the usual method of moments estimation, the LME obtains parameter estimates by equating the first $p$ sample L-moments to the corresponding population quantities. With small and moderate samples, the LME is often more efficient, robust, and computationally convenient than the MLE (Hosking et al.~1985; Naghettini 2017; Hossain et al.~2021). 

\subsection{Redefined return levels}
The $1/p$ return level (RL) is defined as the $1-p$ quantile of the GEVD. For the annual extreme data, sometimes the name ``T-year RL" is also used, with the ``return period" $T=1/p$.
Various return levels are available for the NS model (Salas and Obeysekera 2014). In this study, two types of RLs are considered. The first is the conventional or ``effective” RL (Katz 2013). The conventional $T$-year RL for the GEV model is defined as follows:
\begin{equation} \label{conv_rt}
r^C_T(t)= 
 \mu(t)-\frac{\sigma(t)}{\xi}\Big[1-y_T^{-\xi}\Big],\  \text{for} \ \xi \neq 0,
\end{equation}
\noindent where $y_T =-\log(1-1/T)$. This RL is still useful but may not be sufficient to explain many aspects of the estimation of risk-related quantities for NS extreme data. Thus, the following so-called `redefined RLs' are also considered in this study.

The redefined $T$-year RL ($r^P_T$) by Parey et al.~(2010) is derived by relating the return period to the expected number of events, which is the solution of the following equation:
\begin{equation}\label{Parey_rt}
1 \;=\; \sum_{t=1}^T \;(1- \hat F_t (r^P_T)),
\end{equation}
where $\hat F_t$ denotes the estimated CDF, as in (\ref{cdf-gevd}), of the NS GEV model. The number 1 on the left side of (\ref{Parey_rt}) corresponds to the expected number of events exceeding $r^P_T$ in $T$ years (Salas et al.~2018). %The estimates of the NS GEV parameters are substituted into $F_t$ to solve this equation. 
The redefined RL does not depend on time $t$ but is a constant. %Appendix A briefly describes another redefined RL. 

{In a nonstationary context, no unique definition exists for the RL because extreme values recorded during the time $1,\dots,t$ have a different distribution than the extremes occuring during a different time $1+C,\dots, t+C$ for any $C \ne 0$ (Ribereau et al.~2008). Olsen et al.~(1998) defined another RL based on interpreting the return period as the expected waiting time. Here, return period $T$ or the expected value of the waiting time untill when the flood exceeds $r_T$ for the first time is represented by the following equation (Cooley 2013; Salas et al.~2018):
	\begin{equation}\label{Olsen_rt}
	T \;=\; 1+\sum_{x=1}^\infty \prod_{t=1}^x \;F_t (r_T).
	\end{equation}
	For computational practice, Cooley (2013) and Salas and Obeysekera (2014) recommended to use $x_{max}$ which satisfies $F_t (x_{max})=1$, for the infinite summation. In this study, however, we did not employ this formula due to technical difficulties arising in solving the equation (\ref{Olsen_rt}) related to the summation up to $x_{max}$.}
	%For details and more on the re-defined RLs, see Olsen et al. (1998), Parey et al. (2010), Cooley (2013), Salas and Obeysekera (2014), Salas et al. (2018), and Parey and Hoang (2021).}

\section{L-moment-based algorithm for nonstationary GEV model}
%The existing algorithms and proposed method for the GEV11 model are described first. Then, the algorithms for other models (GEV10 and GEV20) are provided.
 
\subsection{Weighted least squares}  
\label{sec:WLS}
Strupczewski and Kaczmarek (2001) and Strupczewski et al.~(2016) proposed the following WLS method for the GEV11 model:  
%(see  Strupczewski et al.(2016) for the following description):

\begin{itemize}
	\item{}  Step 1. Fit $Z_t = \mu_0 + \mu_1 \times t $ to the data using the least squares method to obtain ($\hat \mu_0,\; \hat \mu_1$) or $\hat \mu_t = \hat\mu_0 + \hat\mu_1 \times t $. 
	\item{}  Step 2. Construct pseudo-residuals as follows: $\epsilon_t = Z_t - \hat \mu_t$.
%	\begin{equation} \label{resid_1}
%	\epsilon_t = Z_t - \hat \mu_t.
%	\end{equation}
	\item{}  Step 3. Construct $\epsilon^\prime _t =| \epsilon_t - \bar{\epsilon} |$, where $\bar{\epsilon}$ is the average of $\epsilon_t$, and implement a regression method to fit $\sigma_t = \text{exp}( \sigma_0 + \sigma_1 t) $ to $\epsilon^\prime _t$ to obtain ($\hat \sigma_0, \;\hat \sigma_1$) or $\hat\sigma_t = \text{exp}(\hat \sigma_0 + \hat \sigma_1 t)$. 
	\item{}  Step 4. Find the WLS estimates ($\hat\mu_0,\; \hat\mu_1$) that minimize  $ \sum_{t=1}^n (Z_t - (\mu_0 + \mu_1 \times t))^2 / {\hat\sigma_t^2} $.
	\item{}  Step 5. Construct $Z_t^\prime = [Z_t - (\hat\mu_0 + \hat\mu_1 \times t)]/{\hat\sigma_t}$.
	\item{}  Step 6. Calculate the stationary LME ($\hat \mu_{st},\ \hat\sigma_{st},\ \hat\xi_{st}$) from $\{ Z_t^\prime \}$ based on the stationary GEVD.
	\item{}  Step 7. Compute a quantile function of the NS model as follows: 
	\begin{equation}\label{SKq}
	X(q, t)= Y(q) \times \hat\sigma_t + (\hat\mu_0 + \hat\mu_1 \times t),
	\end{equation}
where $Y(q)$ is the quantile function (for $0<q<1$) obtained from a stationary GEVD with ($\hat\mu_{st},\ \hat\sigma_{st},\ \hat\xi_{st}$).
\end{itemize}
 
 {The main idea of WLS method is that it fits the data by a NS trend model firstly, and then it applies the stationary LME to the standardized residuals (representing irregular fluctuation around the trend) to obtain the final quantile estimate.  
 	The first part is consisted of using regression model to approximate an initial location parameter, stabilizing the detrended residuals to obtain useful weights and scale parameter, and estimating anew NS location parameter by the weighted regression. In the second part, the stationary LME is computed for the standardized residuals which are assummed to be independent and identically distributed. The final quantile estimates are obtained by combining the above first and scond components.}
 
 Step 7 is sufficient to calculate the conventional RL function. However, for further analysis such as testing hypothesis on a particular parameter or estimating the redefined RL by Parey et al.~(2010), we need to obtain an estimate for each parameter. The details on obtaining them are presented in the Appendix.
 
\subsection{Method by {Gado and Nguyen (2016)} } 
\label{GN16}
For the GEV11 model, the GN16 method applies the following algorithm, where Steps~1 to 3 are the same as those for the WLS method. Step~4 in the WLS method is no longer implemented; instead, the following algorithm is used:

\begin{itemize}
\item{}  Step 5. Construct the following pseudo-residuals using the formula from Cunderlik and Burn (2003):
\begin{equation} \label{Qmax}
Q_t^{max} = \left\{ \begin{array}{ll} &\epsilon_t - sign(\hat\sigma_t) \times \hat\sigma_t  ~~~\mbox{for}~ \epsilon_t \ge \bar \epsilon  \\
     &\epsilon_t + sign(\hat\sigma_t) \times \hat\sigma_t  ~~~\mbox{for}~ \epsilon_t < \bar \epsilon, \end{array} \right.
\end{equation}
where $sign(\hat\sigma_t) = + 1$ or $-1$ accordingly when $\hat\sigma_t$ has an increasing or decreasing trend, respectively.
\item{}  Step 6. Calculate the stationary LMEs ($\hat\mu_{st},\ \hat \sigma_{st},\ \hat \xi_{st}$) from $\{ Q_t^{max} \}$ based on the stationary GEVD.
\item{} Step 7. Estimate the new NS parameters ($\hat\mu_0,\; \hat\mu_1,\; \hat\sigma_0,\; \hat\sigma_1,\; \hat\xi$) through the relationship between the GEV parameters and temporal moments, where $\hat\xi =\hat \xi_{st}$  (detailed in the GN16 method).
\end{itemize}

{The NS parameter estimates by considering the relationship between GEV parameters and the temporal moments in Step~7 are
\begin{eqnarray} \label{time.m}
\hat \mu_t &= \;\hat \mu_0^\prime + \hat \mu_1^\prime \times t - b(\hat \xi) \;\hat \sigma_t, \\
\log (\hat \sigma_t) &=  \hat \sigma_0^\prime + \log \; c(\hat \xi)  + \hat \sigma_1^\prime \times t, 
\end{eqnarray}
where 
\begin{equation}
b(\hat \xi)= \frac{1-\Gamma(1 + \hat\xi_{st})}{\hat\xi_{st}} ~~~ \text{and} ~~~ c(\hat \xi) = \left[\ \frac{\hat \xi_{st}^2}{\Gamma(1+2 \hat\xi_{st})-\Gamma^2 (1+\hat\xi_{st})}\ \right]^{1/2}. \nonumber
\end{equation}
 In (\ref{time.m}), $\hat \mu_0^\prime,\; \hat \mu_1^\prime,\; \hat \sigma_0^\prime,\; \hat \sigma_1^\prime$ are the estimates obtained in the Steps~1 to 3 of the WLS method. It is notable that only $\hat\xi_{st}$ among three stationary LMEs is used in the above formula.}

 {The main idea of the GN16 method is similar to the WLS in the sense that both methods first estimate a trend component and then construct a residuals for which the stationary LME can be applied. 
 	Notably, the formula ($Q_t^{max}$) by Cunderlik and Burn (2003) in Step~5 for constructing pseudo-residuals is different from that ($Z_t^\prime$) of the WLS method. However, the basic idea of constructing $Q_t^{max}$ is similar to that of $Z_t^\prime$ of the WLS method in the sense that $Z_t^\prime$ and $Q_t^{max}$ are the detrended and rescaled series. However, in the Step~7, the NS parameters are estimated differently from the WLS method.}

 Using this method, Gado and Nguyen (2016b) investigated the NS behavior of flood peaks in Quebec. The GN16 method performs better than MLE and stationary LME for most cases, except for the GEV11 model with positive trends in the standard deviation (GN16). Thus, in this study, the attempts to improve the GN16 method are focused on the GEV11 model with positive trends in the standard deviation.
 
 {One reason why the GN16 method did not work well for the case may be related to the stationarity of $Q_t^{max}$, in our opinion. We observed that $\hat \sigma_0^\prime =\hat \sigma_0 + log \;c(\hat \xi)$ in (\ref{time.m}) is sometimes too small, which led to under-estimation of quantiles. This problem may be due to a biased estimation of $\hat\xi_{st}$ from $Q_t^{max}$ which was not well constructed to be stationary. Another reason may be because the parameters are estimated once based on $Q_t^{max}$ and never being updated, without checking the stationarity of $Q_t^{max}$. Thus the proposed method updates some estimates, which were obtained from the GN16 or WLS methods, to make the standardized residuals follow a stationary distribution very well. Another reason is because the GN16 method was heavily influenced by outliers or large values near the end of sample, which were observed in some of our experiments. In the GEV11 model with positive trends in the standard deviation, there was a tendency of showing more outliers toward the end of sample, which sometimes leaded the GN16 method to absurd fit.}
 
\subsection{Proposed method}%: Refinement of Gado and Nguyen method} %: L-moment estimation under a standard Gumbel  distribution}
In the above two methods, the NS sequence was transformed into independent and identically distributed random variables when applying the LME method, as in Step~5 of each method. For this proposal, after taking the similar steps as the Steps~1 and 3 of the WLS method, we considered a different transformation, $\tilde Z_t$, defined as follows: 
\begin{equation} \label{transform}
\tilde Z_t = {{-1} \over {\hat\xi}} \;
\text{log} \left\{ 1 - \hat\xi \;\frac{Z_t -\hat\mu_t}{\hat\sigma_t} \right\}.
\end{equation}
Then, $\tilde Z_t$ follows the standard Gumbel distribution, % denoted by Gumbel(0,1), 
assuming the parameter estimates are true {(Coles 2001, pp.~110).} {See the Supplementary Information for a proof that $\tilde Z_t$ really follows the standard Gumbel distribution.}  $\tilde Z_t$  is referred to as ``standardized residuals.”  %This transformation has been used in model diagnostics and in generating bootstrap samples for fitted NS GEV models (Obeysekera and Salas 2014). 

The proposed method refines the GN16 method as follows:
\begin{itemize}
	\item{Step 1.} Obtain parameters $\hat \mu_0, \; \hat \mu_1,\; \hat \sigma_0,\; \hat \sigma_1,\; \hat \xi$ from the GN16 method, where a robust regression is used in Steps~1 and 3 of the WLS method. % and from (\ref{g0g1}) and (\ref{alpha01}).
	\item{Step 2.} Find $\mu_0, \; \sigma_0$ and $\xi$ that satisfy the following system of three equations under the condition that $\hat \mu_1,\; \hat \sigma_1$ are fixed:
	\begin{eqnarray} \label{LME-NS}
	\lambda_1  = l_1 (\tilde Z_t), ~~~~
	\lambda_2  = l_2 (\tilde Z_t), ~~~~
	\tau_3  = t_3 (\tilde Z_t), 
	\end{eqnarray} 
	where $\lambda_1,\; \lambda_2 $, and $\tau_3$ are the L-moments of a standard Gumbel distribution, and $l_1(\tilde Z_t) ,\; l_2(\tilde Z_t)$, and $ t_3(\tilde Z_t)$ are sample L-moments calculated from the transformed data $\{\tilde Z_t\}$.
\end{itemize}

In Step~1, an ordinary linear model was used in the WLS and GN16 methods for the GEV11 model, whereas Ribereau et al.~(2008) recommended the application of a robust regression method. In addition, GN16 used Sen’s robust estimator for the GEV10 model. Thus, we employed a robust regression using the function lmrob in the `robustbase' package in R (Koller and Stahel 2017).
The robust regression used here is the MM-estimator derived by Yohai (1987), in which the regression parameters are estimated by determining the solution to $p+1$ equations: 
\begin{equation} \label{MM-Reg}
\sum_{i=1}^n \Psi (r_i / \hat \tau) x_{ij} =0,~~~ j=0,\cdots,p,
\end{equation}
where $x_{i0} =1$, $p$ denotes the number of independent variables, and $r_i$ represents the residuals from the ordinary least squared estimation. Here, $\Psi$ is the redescending influence function which down-weights the residuals that are far from zero. Moreover, $\hat \tau$ indicates a robust measure of variation based on the residuals. A popular choice of $\Psi$ is Tukey's biweight function (Wilcox, 2021). A usual selection of $\hat \tau$ is $1.4826\, \{ 1+\, 5 / (n-p) \}  \times \text{med}(|r_i|)$. 

% which is the MM-estimator derived by Yohai (1987).
% in which the regression parameters are estimated by determining the solution to $p+1$ equations: \begin{equation} \label{MM-Reg} \sum_{i=1}^n \Psi (r_i / \hat \tau) x_{ij} =0,~~~ j=0,\cdots,p, \end{equation} where $x_{i0} =1$, $p$ denotes the number of independent variables, and $r_i$ represents the residuals from the ordinary least squared estimation. Further, $\Psi$ indicates the redescending influence function that  down-weights the residuals that are far from zero. Moreover, $\hat \tau$ indicates a robust measure of variation based on the residuals. A popular choice for $\Psi$ is Tukey’s biweight function (Wilcox, 2021). A usual selection of $\hat \tau$ is $1.4826 ( 1+ {5 \over {n-p} } ) \times \text{med}(|r_i|)$. 

 In Step~2, the transformation to $\{\tilde Z_t\}$ is a function of $\mu_0, \; \sigma_0$ and $\xi$; that is,  
 \begin{equation}
 \tilde Z_t =\tilde Z_t (\mu_0,\sigma_0, \xi \;|\; \hat \mu_1,\; \hat \sigma_1 ).
 \end{equation}
 The population $L$-moments ($\lambda_1,\;\lambda_2,\; \tau_3$) are obtained from the standard Gumbel distribution, where $\lambda_1 = 0.572215$ (the Euler constant), $\lambda_2= \text{log}(2)$, and $\tau_3 =0.169925$. 
 
 The proposed method searches the estimates of ($\mu_0,\; \sigma_0$, and $\xi$) which make the standardized residuals $\{\tilde Z_t\}$ follow the standard Gumbel distribution as closely as possible in the L-moment sense. To solve a system of three equations in Step~2, we employed an iterative root-finding routine using the `nleqslv' and `lmomco' packages (Asquith, 2011) in R. {The algorithm in `nleqslv' package is based on Newton's method for nonlinear equations for finding the roots of differentiable
 	functions, that uses iterated local linearization of a function to approximate its roots. This algorithm is powerful and faster in general than other root-finding methods, but convergence is not guaranteed. Convergence depends on the initialization and behaviour of the first derivatives in the neigbohood of a particular root (Dennis and Schnabel 1996). Thus we tried several (such as 20) starting values in our code with a limitted number of iterations. When multiple soultions which satisfying (\ref{LME-NS}) are found, we employed a goodness-of-fit criterion to choose the best among those multiple solutions, which is discussed in Section~\ref{sec:diss}. When a solution satisfying (\ref{LME-NS}) is not found, the proposed method uses the estimates obtained in the previous step. But the failure rate was less than 0.05\% in our experiments.
 }
%Although only two (location and scale) parameters are required in the Gumbel distribution, we consider solving a system of equations with respect to three parameters, including the shape parameter, because the transformed residuals follow the standard Gumbel only when the parameters are valid. 
 %Step~2 resulted in an improvement over the GN16 and other methods.
 
 In Step~2, the proposed algorithm fixes two coefficients ($\hat \mu_1$ and $\hat \sigma_1$) for the time variable and updates the other parameters ($\mu_0,\; \sigma_0$, and $\xi$) that are not directly related to the time variable. The rationale for updating the other parameters is that those parameters may not be obtained well in the GN16 method. 
 This rationale is used again in developing an estimation algorithm for a general NS GEV model with physical covariates, as described in Section~\ref{general_NSGEV}.

\subsection{Methods for GEV10 and GEV20 models} 

Methods for the GEV11 model presented in the previous sections can be similarly applied to the GEV10 and GEV20 models with some modifications. %This section describes such modifications briefly.
 Because the scale parameter is constant in the GEV10 and GEV20 models, we did not require Step~3 in the WLS method. Moreover, $Q_t^{max} = \epsilon_t$ in Step~5 of the GN16 method. %Afterward, we proceeded with the further steps of the GN16 and the proposed method, as in the previous section. For Step~7 in the GN16 method, the NS parameters were estimated using the corresponding relationship between the GEV10 (the GEV20) model and the temporal moments (see the GN16 method for details). 
 For GEV10, Sen’s nonparametric robust estimators for $\mu_0$ and $\mu_1$, as presented in the GN16 method, are employed in Step~1 in the WLS and GN16 methods. For GEV20, the regression coefficients ($\mu_0$, $\mu_1$, and $\mu_2$) are estimated using a robust regression method.

\section{Simulation study} \label{sim_study}
%\subsection{Simulation setting}
\subsection{Simulation setting}
To evaluate the performance of the considered methods, we conducted a small Monte Carlo simulation study where RL is already known. The considered methods for comparison are MLE, stationary LME, WLS, the GN16 method, and the proposed method. We generated 1,000 random samples from the NS GEV11 model. In this simulation, we set $\mu_t$ and $\sigma_t$ as in (\ref{mu_t}) and (\ref{sigma_t}), with sample sizes of $n=25$ and $n=50$. For the parameters of GEV11, we first set $\mu_0 = 0,\; \mu_1 =-.1,\; \sigma_0=1$, and $\sigma_1=.02$, following  GN16, which is a decreasing location and increasing scale parameter setting as time changes. For the shape parameter, which is constant over time, we experimented with the following nine cases (in Hosking-Wallis notation): $\xi =-.35,\; -.25,\; -.15,\; -.05,\;0.0,\; .05,\; .15,\; .25$, and $.35$. {Note again that negative $\xi$ stands for heavy tail, whereas positive $\xi$ represends for light or bounded tail.}
The focus of interest in this study is the estimation of the $T$-year conventional and redefined RLs at the end of the sample. Specifically, we set $T=100$ for the conventional RL  $(r^C_T(t))$ and $T=50$ for the redefined RL $(r^P_T)$. 
%The quantities at time $t$ are represented as follows: \begin{eqnarray} &\text{conventional RL:}~ r_{T}(c;t=n), \label{tend-time}\\
% &r_{T}(c;t=n) \;+\; r_{T}(c;t=n+20) , \label{sumt20}\\  & \text{redefined RL:}~ r_T(P) . \end{eqnarray}
% where $I)$ and $r(P)$ represent the conventional and redefined RLs, respectively.
   
We calculated the following evaluation measures for the above quantities to compare the five estimators: 
\begin{eqnarray} \label{eval}
\text{Bias} = \bar{\hat{r}}_T - r_T , ~~~~
\text{SE} =  [{1\over N}\sum_{i=1}^N (\hat{r}_T(i)- \bar{\hat{r}}_T)^2]^{1/2}, ~~~~
\text{RMSE} = ( Bias^2 + SE^2)^{1/2},
\end{eqnarray}
where $SE$ denotes the standard error, $r_T$ represents the true RL, $\hat{r}_T(i)$ is the RL estimate from the $i$th simulation sample, and $\bar{\hat{r}}_T$ is the average of N number of RLs, where $N$ indicates the number of simulation samples ($N=1,000$). A smaller value is preferred for these measures. 

 \subsection{Simulation results for the GEV11 model} 
Tables~\ref{sim_qt} and \ref{sim_Parey} present the simulation results for the bias, SE, and RMSE for the NS GEV11 model computed using the five methods. The best performance is indicated in bold font. The true RLs at the bottom of Tables~\ref{sim_qt} and \ref{sim_Parey} were calculated from the formulas in (\ref{conv_rt}) and (\ref{Parey_rt}), respectively, using the given parameter setting.

\begin{table}[h!tb]
\begin{center}
	\caption{Simulation results for the conventional 100-year return level at the end of the sample. The bias, standard error (SE), and root mean squared error (RMSE) for the nonstationary generalized extreme value (GEV11) model were computed using five methods, with a sample size of $n=50$, as $\xi$ changes from -.35 to .35.} 
	\label{sim_qt}
	\vspace{0.3cm}
	\begin{tabular}{c|c|ccccccccc}
		\hline
		\multirow{2}{*}{Measure} & \multirow{2}{*}{Method} &  \multicolumn{8}{c}{$\xi$}                                                                                                                     \\ \cline{3-11}
		&                        & -0.35           & -0.25           & -0.15           & -0.05    & 0.0       & 0.05            & 0.15           & 0.25           & 0.35           \\ \hline
		\multirow{5}{*}{Bias}     & MLE                    & 26.71          & 10.12          & 5.87           & 2.09     &1.08      & 0.57           & \textbf{0.24} & -0.75         & -0.77         \\
		& LME               & -30.26         & -22.54         & -16.41         & -12.40   &-10.51      & -9.45          & -7.17         & -5.65         & -4.20         \\
		& WLS                    & \textbf{1.88}  & \textbf{0.12}  & \textbf{0.89}  & \textbf{0.30}  & \textbf{0.17} & \textbf{-0.11} & 0.64          & \textbf{0.17} & \textbf{0.52} \\
		& GN16                   & -9.00          & -10.07         & -7.62          & -6.69     &-6.53     & -6.75          & -5.37         & -5.12         & -4.43        \\
		& PROP                   & -14.41         & -9.83          & -5.19         & -2.90   &-1.13       & -1.19          & 0.48          & 0.56          & 1.11          \\ \hline
		\multirow{5}{*}{SE}     & MLE                    & 185.0         & 51.84          & 26.97          & 17.73     &15.34     & 12.17          & 8.40          & 4.92          & 3.40          \\
		& LME                & \textbf{24.17} & \textbf{16.24} & \textbf{10.46} & \textbf{6.97}  & \textbf{5.25} & \textbf{4.51}  & \textbf{3.04} & \textbf{2.03} & \textbf{1.52} \\
		& WLS                    & 50.35          & 37.80          & 23.47          & 17.27          & 13.33 & 11.24          & 8.43          & 6.44          & 5.25          \\
		& GN16                   & 55.83          & 34.04          & 22.58          & 15.52          & 11.98 &9.30           & 6.85          & 4.94          & 3.72         \\
		& PROP                   & 32.92          & 22.43          & 16.44          & 12.54          & 10.70& 9.49           & 8.03          & 6.76          & 5.77          \\ \hline
		\multirow{5}{*}{RMSE}   & MLE                    & 186.9         & 52.82          & 27.61          & 17.85      & 15.38   & 12.19          & 8.41          & \textbf{4.98} & \textbf{3.49} \\
		& LME               & 38.73          & 27.78          & 19.46          & 14.22          & 11.75&10.47          & \textbf{7.78} & 6.00          & 4.47          \\
		& WLS                    & 50.39          & 37.80          & 23.48          & 17.27          & 13.33&11.24          & 8.46          & 6.45          & 5.27          \\
		& GN16                   & 56.56          & 35.50          & 23.83          & 16.91          & 13.64&11.49          & 8.71          & 7.12          & 5.78          \\
        & PROP        &\textbf{35.93}  & \textbf{24.49}
             &\textbf{17.24}	&\textbf{12.87}  & \textbf{10.76} &\textbf{9.56}	 & 8.04   &  6.78 &  5.88 \\
        \hline
		\multicolumn{2}{c|}{True return level}                         & 79.51          & 58.79          & 43.95          & 33.21     &   28.59  & 25.36          & 19.55         & 15.19         & 11.89        \\ \hline
		\end{tabular}
		\vspace{0.1cm}
	\end{center}
{\small  GN16: method by Gado and Nguyen (2016a), WLS: weighted least squares by Strupczewski and Kaczmarek (2001), LME: stationary L-moment estimation, PROP: proposed method, MLE: maximum likelihood estimation.}
\end{table}

\begin{table}[h!tb]
\centering
	\caption{Same as Table~\ref{sim_qt} but for 50-year return level redefined by Parey et al.~(2010).} 
	\label{sim_Parey}
	\vspace{0.3cm}
	\begin{tabular}{c|c|ccccccccc}
		\hline
		\multirow{2}{*}{Measure} & \multirow{2}{*}{Method} & \multicolumn{8}{c}{$\xi$}                                                                                                                  \\ \cline{3-11}
		&                        & -0.35           & -0.25           & -0.15          & -0.05     &0.0     & 0.05           & 0.15           & 0.25           & 0.35           \\ \hline
		\multirow{5}{*}{Bias}     & MLE                    & 6.56           & 2.73           & 1.62          & 0.57    & \textbf{0.19  }    & \textbf{0.16} & \textbf{0.01} & -0.26         & -0.28         \\
		& LME               & 11.80          & 7.00           & 4.51          & 2.55    &1.72      & 1.33          & 0.66          & \textbf{0.08} & \textbf{0.01} \\
		& WLS                    & \textbf{0.44}  & 0.51           & 0.99          & 0.75    &0.69      & 0.74          & 0.90          & 0.71          & 0.81          \\
		& GN16                   & -2.14        & -3.26         & -3.31         & -3.33         & -2.81 &-3.24         & -2.74         & -2.49         & -1.99         \\
		& PROP                   & 0.45           & \textbf{-0.29} & \textbf{0.04} & \textbf{0.00} & 0.20 & 0.27         & 0.63          & 0.62         & 0.83         \\ \hline
		\multirow{5}{*}{SE}     & MLE                    & 38.84          & 13.83         & 8.71          & 6.02     &5.32     & 4.10          & 2.93          & \textbf{1.88} & \textbf{1.34} \\
		& LME               & 24.17          & 16.24          & 10.46         & 6.97          & 4.90 &4.51          & 3.04          & 2.03          & 1.52         \\
		& WLS                    & \textbf{14.37} & 10.90          & 7.63   &5.89       & 5.66          & 3.93          & 3.04         & 2.47          & 2.08         \\
		& GN16                   & 20.00         & 13.32          & 8.75     &6.01     & 5.14         & 4.06          & 3.13          & 2.35          & 2.21         \\
		& PROP                   & 15.26          & \textbf{10.11} & \textbf{7.07} & \textbf{4.92} &  \textbf{4.24} &\textbf{3.52} & \textbf{2.88} & 2.46          & 2.19         \\ \hline
		\multirow{5}{*}{RMSE}   & MLE                    & 39.39          & 14.10          & 8.86          & 6.05    &  5.32    & 4.11         & \textbf{2.93} & \textbf{1.90} & \textbf{1.37} \\
		& LME               & 26.89          & 17.69          & 11.39         & 7.42         &5.19 &4.70          & 3.11         & 2.03          & 1.51          \\
		& WLS                    & \textbf{14.38} & 10.91         & 7.69          & 5.71         &5.70 &4.00          & 3.18         & 2.57         & 2.24          \\
		& GN16                   & 20.11          & 13.72          & 9.36         & 6.92          &5.86 &5.20          & 4.16          & 3.42         & 2.91          \\
		& PROP                   & 15.27          & \textbf{10.11} & \textbf{7.07} & \textbf{4.92} & \textbf{4.25} &\textbf{3.53} & 2.94          & 2.53          & 2.34          \\ \hline
		\multicolumn{2}{c|}{True return level}                         & 37.44          & 29.24          & 23.02         & 18.25     &  16.47  & 14.58         & 11.71         & 9.46          & 7.66              \\ \hline
		
	\end{tabular}
\end{table}

\begin{figure}[!htb]
	\centering
	\begin{tabular}{l}	\includegraphics[width=12cm, height=12cm]{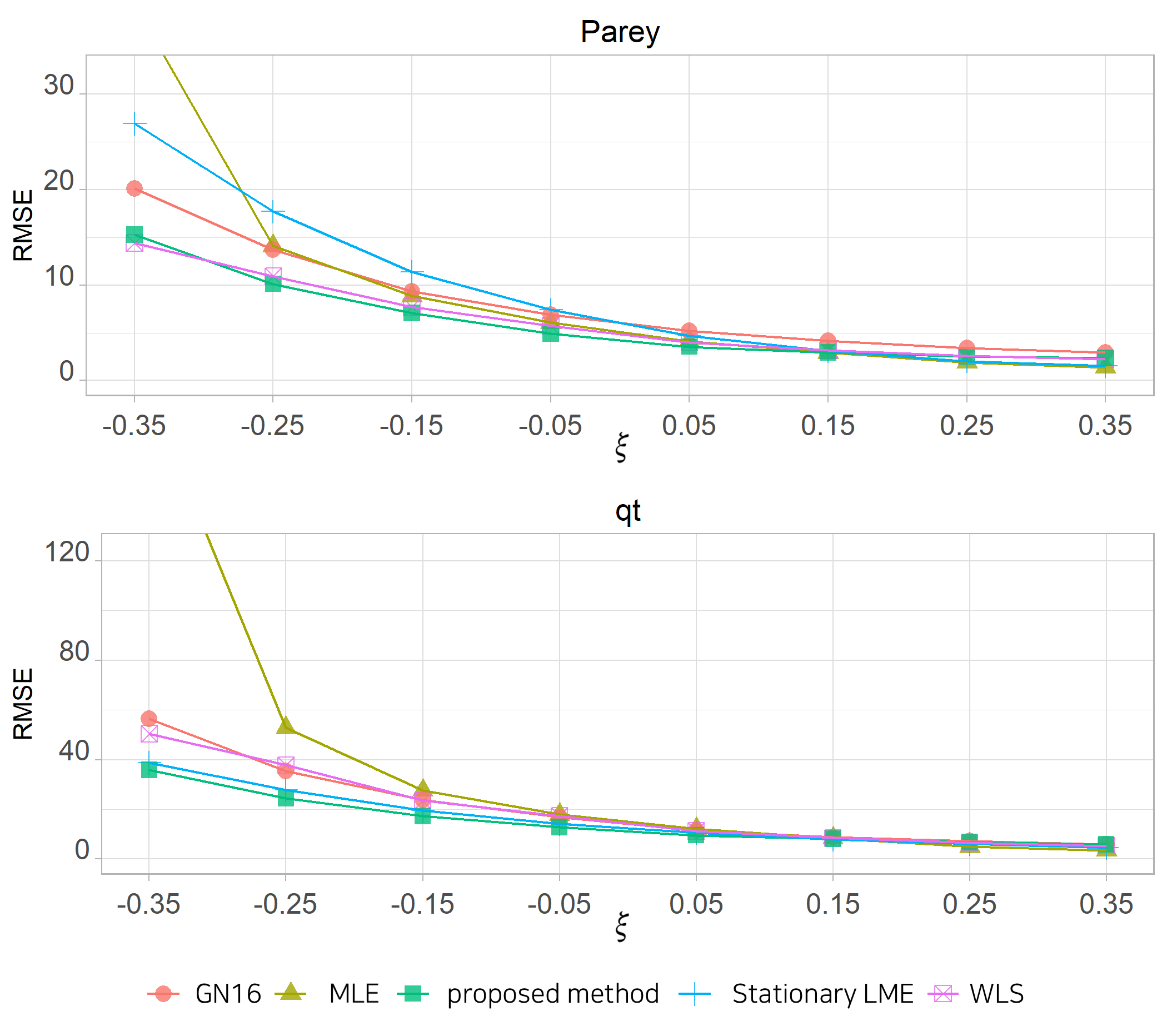} \end{tabular}	
	\caption{Simulation results: Root mean squared error (RMSE) of the 50-year return level (RL) redefined by Parey et al.~(2010), the conventional 100-year RLs at the end of the sample ($qt$) from the GEV11 model fitted using the five methods, with a sample size of $n=50$, as $\xi$ changes from -.35 to .35. GN16: method by Gado and Nguyen (2016a), WLS: weighted least squares,
		% LME.STA: stationary L-moment estimation,
		 MLE: maximum likelihood estimation.} \label{fig:sim-1}
\end{figure}

The proposed method performed best for $\xi \le .15$ in terms of the RMSE. The MLE worked well for $\xi \ge .2$ in terms of the RMSE but worked poorly for $\xi \le -.25$. The differences in the RMSE values among methods for $\xi \ge .15$ were relatively minimal compared to the actual values at the bottom of each table (see Figure~\ref{fig:sim-1} too). Notably, in Table~\ref{sim_qt}, for the conventional RLs, the stationary LME has the lowest SE but the highest bias. In Table~\ref{sim_Parey}, for the redefined RL, the stationary LME had high values for the bias and SE. For the redefined RL, the WLS worked well for $\xi \le -.30$. Table~\ref{sim_qt} reveals that the WLS method has a low bias but a high SE.% and exhibits a higher RMSE value than the stationary LME.

The Supplementary Information provides 40 sample plots of 50-year conventional RL estimates obtained using the five methods for the GEV11 model. The random samples were generated under the settings for which $\mu_0 = 0,\; \mu_1 =-.1,\; \sigma_0=1,\; \sigma_1=.02 $, and $\xi = -.2$. From these plots, with $n=50$, we observe that the behaviors of the MLE, WLS, and proposed methods are similar to some degree. The proposed method rarely results in RL plots far from the true RL plot. MLE works well in general, except in a few cases where the MLE exhibits an extremely high bias at the end of a sample. %resulting in a high RMSE for $\xi \le -.2$ in the simulation study. 
The performance of WLS is between that of the proposed method and MLE. %, whereas WLS exhibits a high bias at the end of a sample in a few cases. %The GN16 method sometimes significantly underestimates or overestimates the true function.

The RL plots of the MLE or WLS methods for the GEV11 model occasionally display parabolic curves far from the true RL function, reflecting that outliers or large values toward the end of a sample significantly influence these methods. For the same data, the proposed method often exhibits the lowest absolute slope line. Thus, the proposed method for the GEV11 model may be resistant to influential observations or outliers.

{The efficiency of the proposed method may be due to the effectiveness of LME and because the transformed values follow the standard Gumbel distribution as closely as possible. Although the WLS and GN16 methods also employ LME, the proposed method goes further (after the GN16 method) by computing LME one more time in Step~2, minimizing the discrepancy in the sense of L-moments between the standardized residuals and standard Gumbel distribution. This additional computation may lead to a more accurate estimation. In the GEV11 model with positive trends in the standard deviation, there are a tendency of showing more outliers toward the end of sample. We believe that employing the robust regression in Step~1 and double usage of L-moments in the proposed method may prevent the fit being too sensitive to outliers.}

%The Supplementary Information presents additional tables and figures containing the results of the simulation with a different setting for the GEV11 model, where $\mu_0 = 0,\; \mu_1 =.2,\; \sigma_0=1$, and $\sigma_1=.02 $, following the GN16 method. In this setting, the location and scale parameters increase over time. The results of this simulation are similar to the results for the first setting above.

\subsection{Simulation of the GEV10 and GEV20 models} 
For the parameters of GEV10, we set $\mu_0 = 0,\; \mu_1 =-.2$, and $\sigma= 1$, following GN16. For the parameters of GEV20, we set $\mu_0 = 50,\; \mu_1 =-.02,\; \mu_2 =.06$, and $\sigma= 33$. 
%, which are similar to the parameter estimates for the peak streamflow data in Trehafod, UK, in the next section.

\begin{table}[h!tb]
	\centering
	\caption{Simulation results for the root mean squared error of the conventional return levels at the end of the sample for the nonstationary GEV10 and GEV20 models with sample sizes of $n=25$ and $n=50$.} 
	\label{tab:GEV10_20_rtn}
	\vspace{0.4cm}
	\begin{tabular}{c|c|c|ccccccccc}
		\hline
		\multirow{2}{*}{Model}  & \multirow{2}{*}{$n$}  & \multirow{2}{*}{Method} & \multicolumn{8}{c}{$\xi$}                                    \\ \cline{4-12}
		&                     &                         & -0.35 & -0.25 & -0.15 & -0.05 &0.0 & 0.05 & 0.15 & 0.25 & 0.35 \\ \hline
		\multirow{10}{*}{GEV10} & \multirow{5}{*}{25} & MLE   & 12.1  & 9.8   & 6.7   & 5.5 &4.6  & 3.9  & 3.4  & 2.7  & 1.9  \\
		&                     & LME                & \textbf{11.7}  & \textbf{8.3}   & \textbf{5.3}  & \textbf{4.6} &4.4 & 4.0  & 3.9  & 3.9  & 3.8  \\
		&                     & WLS                     & 12.6  & 9.1   & 5.9   & 4.8 &4.1  & 3.6  & 2.9 & 2.4  & 1.7  \\
		&                     & GN16                    & 12.7  & 10.0  & 7.6   & 6.0 &5.0  & 4.6  & 3.9  & 3.5  & 2.8  \\
		&                     & PROP                    & 12.7  & 9.2   & 6.4   & 5.0 &\textbf{4.0 } & \textbf{3.4}  & \textbf{2.6}  & \textbf{2.2}  & \textbf{1.5}  \\ \cline{2-12}
		& \multirow{5}{*}{50} & MLE                     & 10.0   & 7.1  & 5.6   & 4.1 &3.5  & 2.9  & \textbf{1.8}  & \textbf{1.3} & \textbf{1.0}  \\
		&                     & LME                & \textbf{9.6}   & 7.1   & 6.3   & 5.9 &6.1   & 6.0  & 6.1  & 6.7  & 7.3  \\
		&                     & WLS                     & 10.5  & 7.1   & \textbf{4.9}   & \textbf{3.4} & \textbf{3.0}& 2.7  & 1.9 & 1.4  & 1.1  \\
		&                     & GN16                    & 10.5  & 9.0  & 6.9   & 5.8  &4.9 & 4.6  & 3.6  & 3.1  & 2.7  \\
		&                     & PROP                    & 9.7   & \textbf{6.9}   & 5.3   & 3.7   & 3.3& \textbf{2.6}  & \textbf{1.8}  & 1.4  & 1.1 \\ \hline
		\multirow{10}{*}{GEV20} & \multirow{5}{*}{25} & MLE  & 22.5 & 18.4  & 16.8  & 13.7 &10.9 & 9.6  & 8.3  & 6.6  & 5.5  \\
		&                     & LME               & 25.3  & 18.0  & 13.1  & 9.8 &8.6  & 7.7  & 6.8  & 6.7  & 6.3  \\
		&                     & WLS                     & 20.6  & 15.7 & \textbf{11.2}  & \textbf{8.0}  &7.4 & \textbf{6.0}  & 5.0  & 3.9  & 3.4  \\
		&                     & GN16                    & 21.2  & 16.7  & 13.0  & 9.5 &7.9 & 7.5  & 6.3  & 6.0  & 5.3  \\
		&                     & PROP                    &  \textbf{19.3}  &  \textbf{15.1}  & 12.2  & 8.3 &\textbf{7.2} & 6.3  & \textbf{4.8} & \textbf{3.8} & \textbf{3.3 } \\ \cline{2-12}
		& \multirow{5}{*}{50} & MLE                     & 21.7  & 17.6  & 16.0  & 13.2 &10.8 & 9.4  & 8.0  & 6.4  & 5.1  \\
		&                     & LME                 & 20.7  & 18.0  & 13.0  & 9.5 &8.7  & 7.5 & 6.4 & 6.3 & 5.9 \\
		&                     & WLS                     & 18.8  & \textbf{13.7}  & \textbf{8.6}   & \textbf{5.9} &\textbf{5.4}  & \textbf{4.5} & 3.8  & 2.6  & \textbf{2.3}  \\
		&                     & GN16                    & 20.0  & 16.5  & 11.5  & 8.7 &7.6  & 7.0  & 6.2  & 5.0  & 4.6  \\
		&                     & PROP                    & \textbf{18.5}  & 15.6  & 9.3   & 6.2 &5.5  & 4.7  & \textbf{3.6}  & \textbf{2.5}  & 2.4  \\ \hline
		
	\end{tabular}
\end{table}

Simulation results for the GEV10 and GEV20 models for the sample sizes of $n=25$ and $n=50$ are presented in Table~\ref{tab:GEV10_20_rtn} for 100-year the conventional RLs at the end of the sample. These tables reveal that % and Figure \ref{fig:GEV10} for the GEV10 model. 
	stationary LME often performs the worst among the five methods, except for the GEV10 model with $n=25$. The GN16 method outperforms MLE and stationary LME for GEV20, consistent with the results of GN16. The proposed method and WLS perform similarly well and sometimes share the best rank.  
	
	{For cases when $\xi$ positive (light or bounded tail), for the GEV11, GEV10, and GEV20 models, it is hard to conclude which method performs the best; the MLE or WLS or proposed method work well for some cases. But it is notable that the absolute values of the RMSE for $\xi$ postive are relatively smaller than those for $\xi$ negative, as seen in Figure \ref{fig:sim-1}. Thus the choice of estimation method for $\xi$ postive may be less critical than for $\xi$ negative.}
	
%Compared to the results for the GEV11 model, the superiority of the proposed method to the WLS and MLE method may be more effective when time dependency in the scale parameter occurs. 
%The Supplementary Information presents more tables and figures for the simulation results. 

\section{Application: Peak streamflow in Trehafod}

For an application, we considered the daily observed extreme flow data from the Rhondda River (Station ID:57006) in Trehafod in the UK. The dataset consists of peak flows from 1968 to 2021, spanning 54 years, with units in cubic meters per second (m$^3$/s).

\begin{figure}[!htb]
	\centering
	\begin{tabular}{l}	\includegraphics[width=12cm, height=10cm]{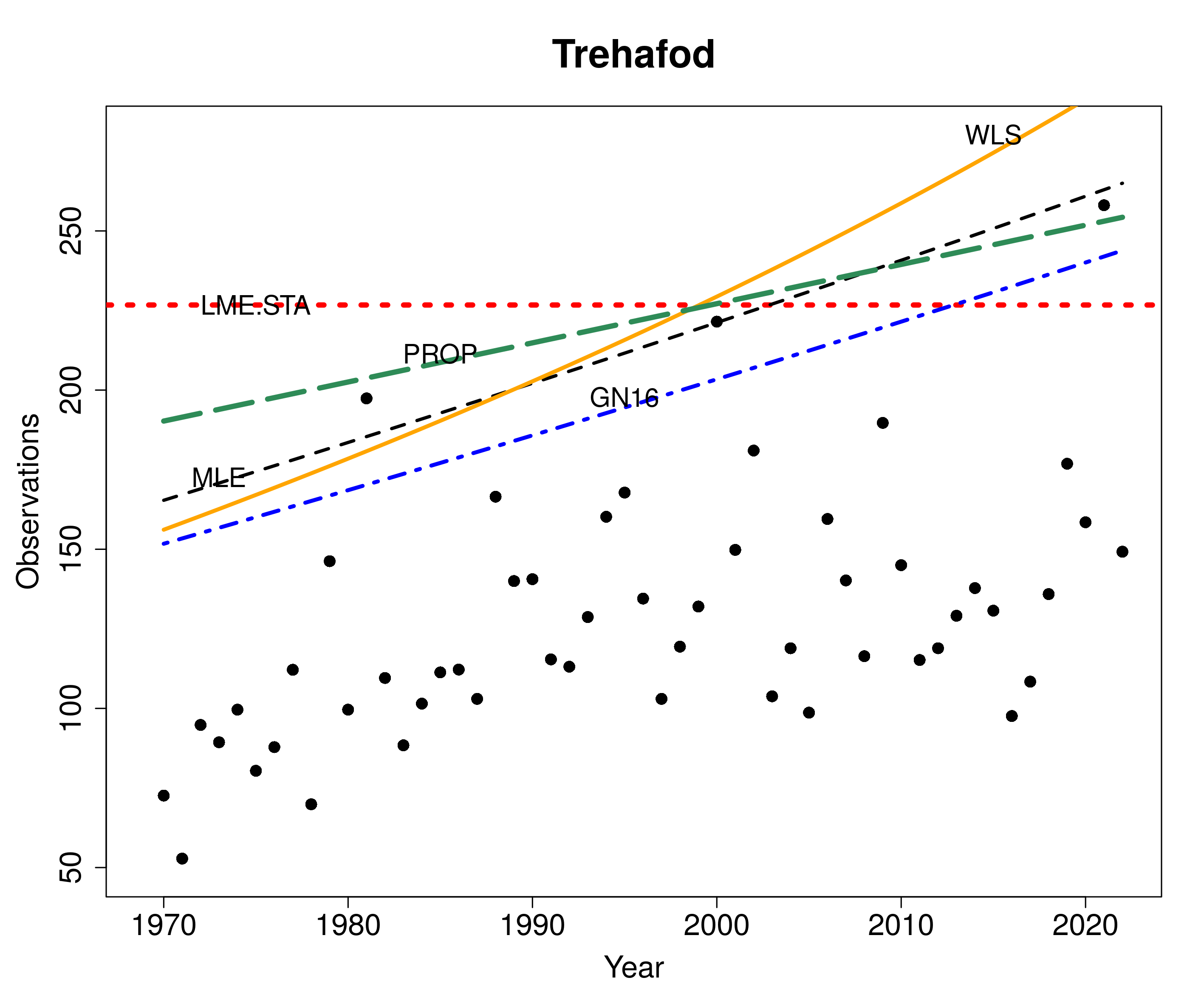} \end{tabular}	
	\caption{Scatterplot and 50-year conventional return levels for the GEV11 model fitted using several methods for extreme flow data (unit: m$^3$/s ) in Trehafod, UK. GN16: method by Gado and Nguyen (2016a), PROP: proposed method, WLS: weighted least squares, LME.STA: stationary L-moment estimation, MLE: maximum likelihood estimation.} 
\label{rt-plot-trehafod}
\end{figure}

Figure~\ref{rt-plot-trehafod} displays a scatterplot of the data and the 50-year conventional RLs obtained from the GEV11 model fitted by the five methods. %An increasing trend was observed from a scatterplot of the data. To assess  this trend, we executed the Mann--Kendall trend test. As a result, the null hypothesis (no linear trend) was rejected at the 5\% level. 
Among the four fitting methods, the WLS exhibited the steepest slope, whereas the proposed method had the least steep slope. 
Table~\ref{tab:trehafod} lists the parameter estimates and 50-year RL estimates. %Kolmogorov--Smirnov (KS) statistics, and their $p$-values from the GEV11 model. %The RL estimates are obtained using the conventional method ($r^C_T (t=n) $) and the redefined method by Parey et al.~(2010) ($r^P_T $). %We did not aim to fit the NS Gumbel models in this real-world data application but focused instead on the GEV11 model.

\begin{table}[!htb]
  \begin{center}
	\caption{Parameter estimates and 50-year return level estimates obtained from the GEV11 model using several methods for the extreme streamflow data in Trehafod, UK (unit: m$^3$/s).} \label{tab:trehafod}
	\vspace{0.3cm}
	\begin{tabular}{|c|c|c|c|c|c|c|c|}
		\hline
		Method & $\hat \mu_0$ & $\hat\mu_1$ & $\hat\sigma_0$ & $\hat\sigma_1$ & $\hat\xi$ & $r^P_T $  &
		$r^C_T (n)$   \\ \hline
		Gum.LME.STA & 110.7 & ~ & 30.15 & ~ & ~ & 228.4 & 228.4  \\ \hline
		GEV.LME.STA & 110.9 & ~ & 30.57 & ~ & .015 & 226.7 & 226.7 \\ \hline
		MLE & {81.3} & {1.19} & {2.99} & .008 & {-.052} & {226.2} & 269.1 \\ \hline
	%	WLS.STA& -.7    &~ & 1.7  & ~ &  -.063& ~& & & \\
		WLS & 91.8 & 1.12 & 2.64 & .016 & -.063 &  {233.9} & 298.2
		\\ \hline
		PROP & 82.9 & 1.09 & 3.12 & .0013 & -.093 & {225.1} & 254.3 \\ \hline
		GN16 & 89.6 & 1.20 & 2.64 & .008 & -.050 & {205.8} & 238.1 \\ \hline
	\end{tabular}
\end{center}
	\vspace{0.1cm}
 \begin{small} STA: stationary; $r^P_T $ and $r^C_T (n)$ represent the return level estimates for $T=50$ computed using the formula proposed by Parey et al.~(2010) and the conventional method at $t=n$, respectively.\end{small}
\end{table} 

\begin{figure}[!htb]
	\centering
	\begin{tabular}{l}	\includegraphics[width=14cm, height=10cm]{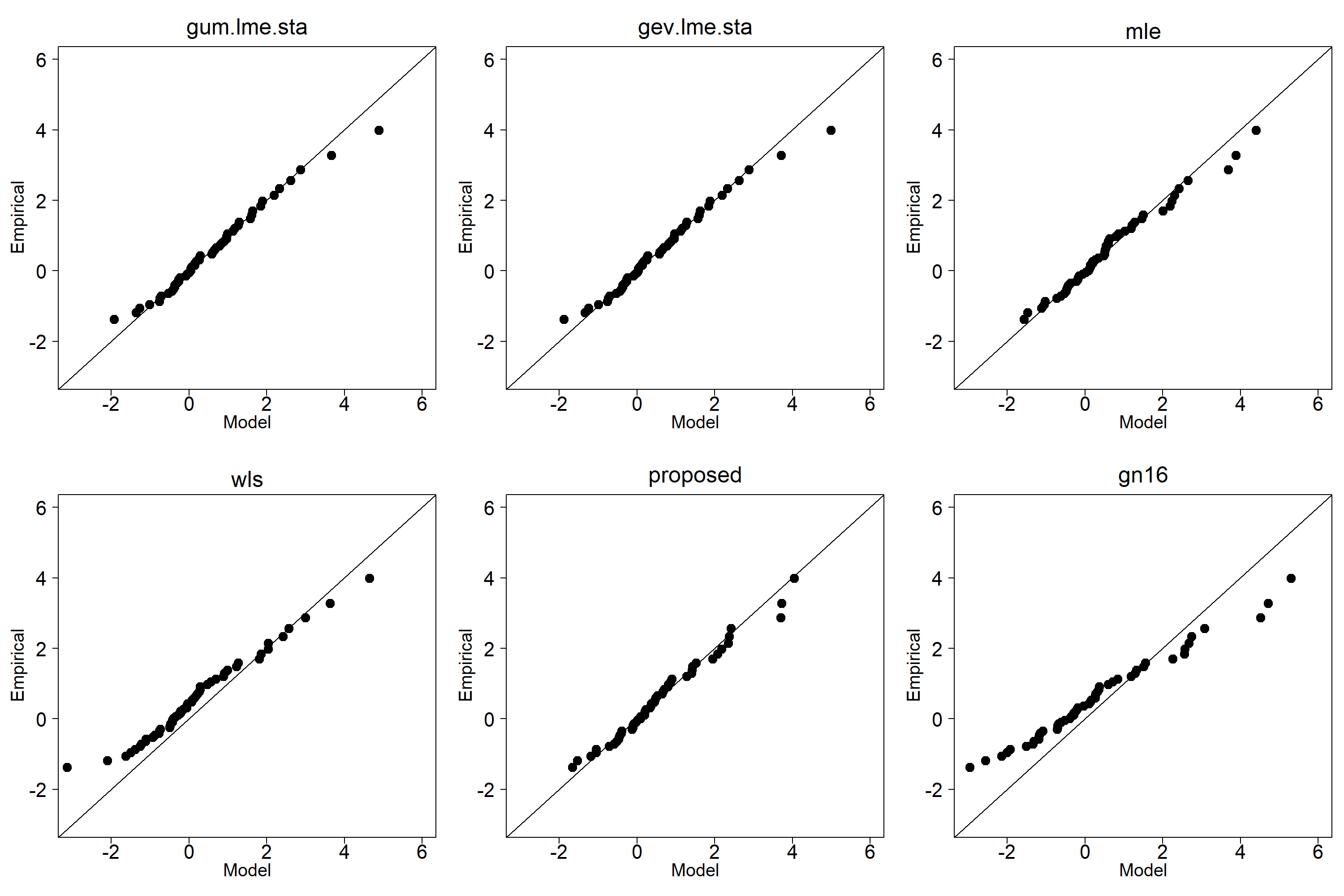} \end{tabular}	
	\caption{Quantile-per-quantile plots for standardized residuals based on the five methods, where the GEV11 model was fitted to the extreme streamflow data in Trehafod in the UK (unit: m$^3$/s).} \label{qqplot}
\end{figure}

Figure~\ref{qqplot} illustrates the quantile-per-quantile (Q-Q) plots for the standardized residuals based on the five methods obtained from the GEV11 model. Two Q-Q plots for the stationary LME methods are remarkably similar because those parameter estimates are almost equal, as listed in Table~\ref{tab:trehafod}. The Q-Q plots of the stationary LME method indicate that a discrepancy occurs in the upper end (one or two points). This discrepancy, especially at the upper end, reveals a significantly biased estimation of the high quantiles. The Q-Q plots for the proposed and MLE methods look better than those for other methods.

\begin{table}[h!bt]
	\centering
	\caption{$T$-year return levels obtained using the formula by Parey et al.~(2010) for several methods, where the GEV11 model was fit to the extreme streamflow data for Trehafod, UK (unit: m$^3$/s). } \label{tab:Parey}
	\vspace{0.3cm}
	\begin{tabular}{c|cccccc}
		\hline
		\multirow{2}{*}{Method} & \multicolumn{6}{c}{Return level ($T$-year)}                 \\ \cline{2-7}
		& 2     & 10    & 20    & 50    & 100   & 200          \\ \hline
		Gum.LME.STA             & 121.8 & 178.6 & 200.3 & 228.4 & 249.4 & 270.4     \\
		GEV.LME.STA             & 122.1 & 178.6 & 199.7 & 226.7 & 246.7 & 266.4      \\
		MLE                     & 90.5  & 137.4 & 164.2 & 226.2 & 334.5 & 630.6     \\
		WLS                     & 98.8  & 135.6 & 160.5 & 233.9 & 416.1 & 1471.9    \\
		PROP                    & 92.9  & 146.0 & 173.4 & 225.1 & 293.1 & 422.5     \\
		GN16                    & 96.6  & 131.5 & 152.8 & 205.8 & 299.9 & 549.1      \\ \hline
	\end{tabular}
        \vspace{0.5cm}
        
{\small Gum.LME.STA: stationary Gumbel distribution L-moment estimation, GEV.LME.STA: stationary generalized extreme value distribution L-moment estimation, MLE: maximum likelihood estimation, WLS: weighted least squares, PROP: proposed method, GN16: method by Gado and Nguyen (2016a).}
 
\end{table}

\begin{figure}[h!bt]
	\centering
	\begin{tabular}{l}	\includegraphics[width=12cm, height=9.5cm]{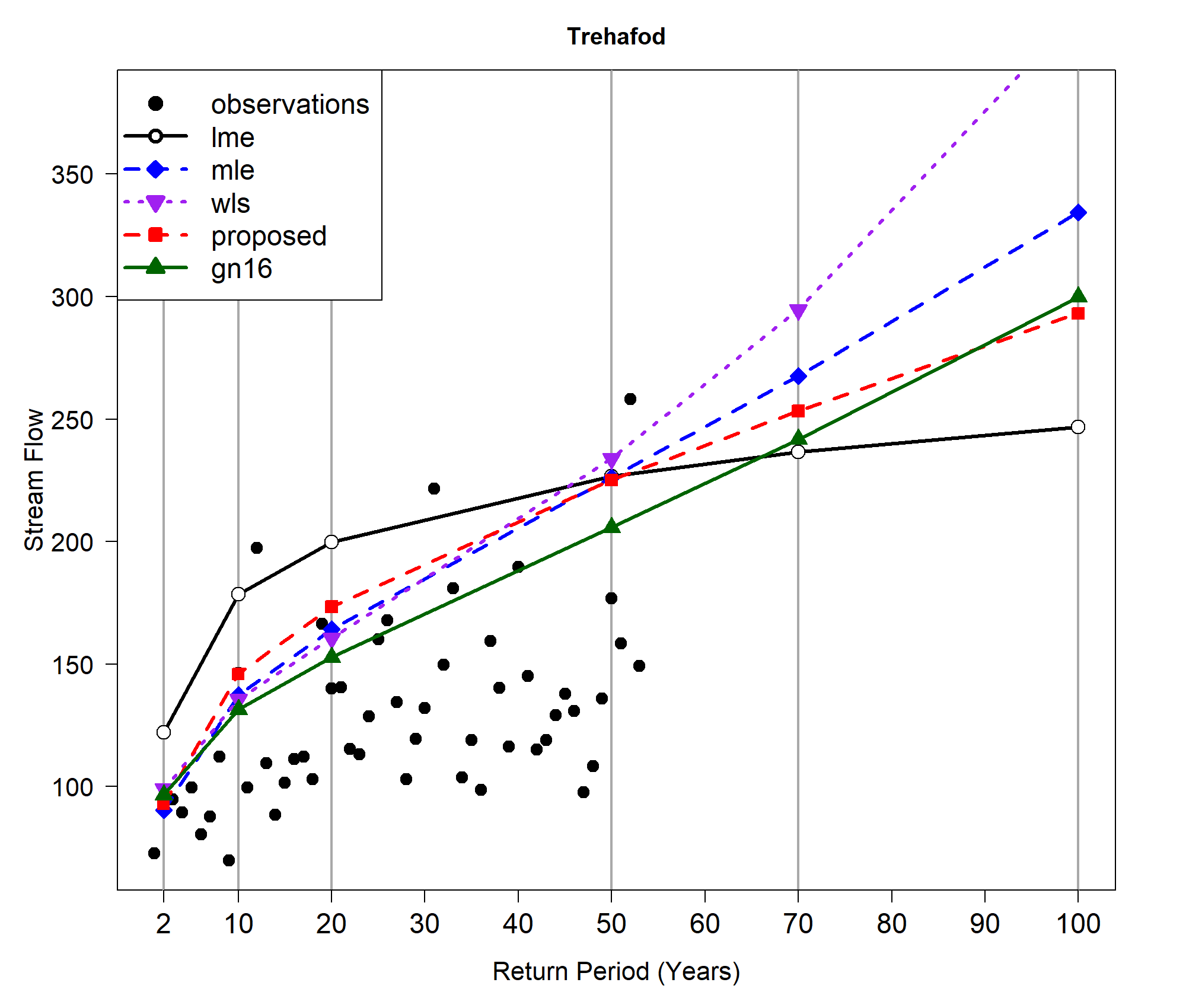} \end{tabular}	
	\caption{Modified return level plots based on the formula by Parey et al.~(2010) for the five methods, where the GEV11 model was fit to the extreme streamflow data for Trehafod, UK (unit: m$^3$/s).} \label{Parey}
\end{figure}

Table~\ref{tab:Parey} and Figure~\ref{Parey} present the RL estimates and their plots based on the formula by Parey et al.~(2010) obtained from the GEV11 model for the five methods. The RL plot in Figure~\ref{Parey} is a modified version of the usual RL plot in Coles (2001). This version is convenient to observe how the RL associated with the observations behaves as the return period changes up to 100 years. The four methods (MLE, stationary LME, WLS, and the proposed method) have similar estimates at $T=50$. For a return period longer than $T=50$, the WLS provides the highest estimates, whereas the stationary LME provides the lowest estimates. For the stationary LME, it is unusual for the RL line to increase slowly for return periods longer than $T=50$. %One observation is still over the 100-year RL, illustrating that the stationary LME significantly underestimates the high quantiles and may result in serious bias. However, the SE may be low, as observed in the simulation study.

\section{Extension and model selection based on L-moments}

\subsection{Extension to nonstationary model with physical covariates}
\label{general_NSGEV}

One limitation of the proposed method is that only the time variable was employed as the covariate of the NS parameters. In this section, we extend the proposed method when covariates other than the time variable are included in the NS GEV model.  

The NS GEV parameters can be written in the form: for example,
\begin{eqnarray} %\label{gen_mu_sigma}
\mu(X)& =& \mu_0 + \mu_1 X_1 +\cdots + \mu_k X_k \label{gen_mu2} \\
\sigma(X)& =& \text{exp}\left( \sigma_0 + \sigma_1 X_1 +\cdots + \sigma_k X_k   \right), \label{gen_sigma2}
\end{eqnarray}
where $X$ indicates a covariate vector and $k$ is the number of covariates.

The central feature of the proposed method in the above sections was two steps. The first step was to estimate all parameters using a regression (the least squares or robust) method. The second step was to estimate the intercepts ($\mu_0$ and $\sigma_0$) and the shape parameter $\xi$ using stationary LME, fixing the estimates of other parameters. Applying this main feature similarly, we extend the proposed method to the NS GEV model in which physical covariates are employed as predictors. For an NS GEV model with $(\mu(X),\;\sigma(X),\; \xi )$, our extension is as follows:

\begin{itemize}
	\item{ Step 1.} Fit $Z_X = \mu_0 + \mu_1 X_1 +\cdots + \mu_k X_k  $ to the data using robust regression to obtain ($\hat \mu_0, \hat \mu_1,\cdots,\hat \mu_k $). 
	\item{Step 2.} Construct absolute pseudo-residuals using $\epsilon_X = | Z_X - \hat \mu(X)|$.
	\item{Step 3.} Fit $\epsilon_X = \text{exp}\left( \sigma_0 + \sigma_1 X_1 +\cdots + \sigma_k X_k \right)$ using regression to obtain ($\hat \sigma_0, \hat \sigma_1,\cdots,\hat \sigma_k$).
	\item{Step 4.} Find $\mu_0, \; \sigma_0$ and $\xi$ that satisfy the following system of three equations under the condition that $(\hat \mu_1,\cdots,\hat \mu_k)$ and $( \hat \sigma_1,\cdots,\hat\sigma_k)$ are fixed:
	\begin{eqnarray} \label{gen-LME-NS}
	\lambda_1  = l_1 (\tilde Z_X), ~~~
	\lambda_2  = l_2 (\tilde Z_X),~~~
	\tau_3  = t_3 (\tilde Z_X), 
	\end{eqnarray} 
	where $\lambda_1,\; \lambda_2 $, and $\tau_3$ are the L-moments of a standard Gumbel distribution, and $l_1(\tilde Z_X) ,\; l_2(\tilde Z_X)$, and $t_3(\tilde Z_X)$ are sample L-moments calculated from the transformed data $\{\tilde Z_X\}$. 
\end{itemize}

In Steps~1 to 3 of the above algorithm, we followed the WLS method by Strupczewski and Kaczmarek (2001) and and Strupczewski et al.~(2016), except for the robust regression in Step~1. %The parameter adjustment in the GN16 method using the relationship between the GEV parameters and temporal moments was not applied in this algorithm. 
The pseudo-residuals $Q_t^{max} $ in Step~5 of the GN16 method are no longer employed. The transformation to $\{\tilde Z_X\}$ in Step~4 is a function of $\mu_0, \; \sigma_0$ and $\xi$; that is,
$~\tilde Z_X = \tilde Z_X\, \{\mu_0,\, \sigma_0,\, \xi \; |\; (\hat \mu_1,\cdots,\hat \mu_k), \, ( \hat \sigma_1,\cdots,\hat\sigma_k) \}$.
To solve a system of three equations in Step~4, we employed an iterative root-finding routine using the `nleqslv' package (Dennis and Schnabel 1996) in R. 

\subsection{Model selection using cross-validated L-moment distance}

Because we use L-moments in fitting NS GEV models, employing the likelihood-based criteria for the model selection does not seem appropriate. Thus, we consider a L-moment-based criterion for model selection, which is a cross-validated (CV) generalized L-moment distance (GLD).

For a given GEV model with parameter estimates, data can be transformed to $\{\tilde Z_X\}$ which may follow a standard Gumbel. Let denote 
\begin{equation} 
\underline{\bf \lambda} =(\lambda_1,\; \lambda_2,\; \tau_3,\; \tau_4)^t ~~\text{and} ~~
\underline{\bf l} =(l_1(\tilde Z_X),\; l_2(\tilde Z_X),\; t_3(\tilde Z_X) ,\; t_4(\tilde Z_X))^t
\end{equation}
are vectors of the population L-moments of a standard Gumbel and sample L-moments of $\{\tilde Z_X\}$.
The GLD is defined as
\begin{equation} \label{GLD}
GLD \ = \ (\underline{\bf \lambda} -\underline{\bf l})^t \ V^{-1}  (\underline{\bf \lambda} -\underline{\bf l}) ,
\end{equation}
where $V$ is the $4 \times 4$ variance-covariance matrix of the sample L-moments of $\{\tilde Z_X\}$. Here, $V$ can be calculated using the formula by Elamir and Seheult (2004). The GLD has been used by Kjeldsen and Prosdocimi (2015) and Shin et al.~(2025) 
~for regional frequency analysis and model averaging, %and generalized L-moment estimation, 
respectively. %We considered L-moments up to 5th order in this study, but one can use it up to 3rd or 4th order.

For model selection purpose, the CV criteria have been widely employed (e.g., Smyth 2001; James et al.~2021; Stein 2021; Fauer and Rust 2023), because it is suitable to find a good model preventing the overfitting. Thus, we also applied the cross-validation idea to the GLD, for selecting the best model among the candidate NS GEV models which were built by the L-moment method. Because L-moments are less sensitive to outliers, we expect that the CV GLD is robust to outliers. To the best of our knowledge, the CV GLD has not been considered in the literature yet. %The CV GLD is applied in the next real data example.

\subsection{Application to maximum sea-levels in Fremantle}
%\subsection{Application to maximum sea levels in Fremantle}
For a real data application of the NS GEV model with a physical covariate, we considered the annual maximum sea level (unit: meters) series recorded at Fremantle, West Australia. These data were analyzed by Coles (2001) and available from the `ismev' package of R. The dataset consists of 86 observations from 1897 to 1989, spanning 93 years, with some missing data.

Coles (2001) fitted three NS GEV models with the following location parameters to these data using MLE with the Southern Oscillation Index (SOI) and time variables, with constant scale and shape parameters:
	\begin{eqnarray} \label{ns_model_fremantle}
\text{Model} ~1: ~ \mu (X)& =& \mu_0 + \mu_1 \times t , \nonumber \\
\text{Model} ~2: ~ \mu (X)& =& \mu_0 + \mu_1 \times SOI(t) , \\
\text{Model} ~3: ~ \mu (X)& =& \mu_0 + \mu_1 \times t + \mu_2 \times SOI(t), \nonumber
\end{eqnarray}
where $t$ denotes the index for the year, with $t=1$ corresponding to 1897. Even though there are the missing data, $t$ spans from 1 to 93. %Based on the likelihood ratio test, Model 3 was selected as an appropriate model by Coles (2001).

We fitted the above three models to the Fremantle data using our algorithm. Table~\ref{tab:app_frem} provides the parameter estimates with SEs in parenthesis and the CV GLD values. The LME result from the stationary GEV model (Model 0) and MLE results from the Model 1 provided in Coles (2001) are presented for comparison. We applied the parametric bootstrap technique to compute the SEs of parameter estimates, {as following;
\begin{itemize}
	\item{Step 1.} Generate $B$ bootstrap samples from the standard Gumbel distribution. Denote $\tilde Z_X$ be the bootstrap sample.
	\item{Step 2.} Transform $\tilde Z_X$ to obtain $Z_X$ using 
	\begin{equation}
	 Z_X = \frac{\hat\sigma(X)}{-\hat\xi(X)} \exp \{-\hat\xi(X)\ \tilde Z_X \} \ + \ \hat\mu(X),
	 \end{equation}
	 which is the back-transformation of (\ref{transform}).
	 \item{Step 3.} Fit the NS GEV model for $Z_X$ using considered methods. 
	 \item{Step 4.} Repeat Steps 2 and 3 for $B$ bootstrap samples to compute the SEs of parameter estimates.
\end{itemize}
In this study, $B$ was set to 300.}

For calculating CV GLD, we used 20 times 5-fold cross-validation. The training and test data are randomly selected and fixed for each time. The covariance matrix was obtained after pooling the transformed full data from four models (Model 0 to 3) and was fixed throughout all the CV GLD computation.

In Table~\ref{tab:app_frem}, the estimates and SEs in Model 1 are similar to those of the MLEs. {Standard errors of $\hat \mu_1$ for the time $t$ are smaller than those for variable SOI. This may be mainly due to the unit difference between the time $t$ and SOI. The coefficients of variation (SE/$\hat \mu_1$) of $\hat \mu_1$ for the time $t$ are 0.316, 0.25, and 0.3 for Model 1, Model 1*, and Model 3, respectively. Whereas the coefficients of variation of $\hat \mu_1$ for SOI are 0.367 and 0.328 for Model 2 and Model 3, respectively. The differences between coefficients of variation for the time $t$ and SOI are not so remarkable. The similar argument using the coefficient of variation can be applied for comparison to SEs of other parameters.}
Finally, the Model 3 is selected as an appropriate model because of the smallest CV GLD, which is consistent with the model selection based on the likelihood tests in Coles (2001).

\begin{table}[h!bt]
	\begin{center}
		\caption{The parameter estimates with standard errors in parenthesis and the cross-validated generalized L-moment distance (CV GLD) values for four models obtained by the L-moment-based method.} \label{tab:app_frem}
		\begin{tabular}{l|ccccc|c}
			\hline
			Model & $\hat \mu_0$& $\hat \mu_1$ & $\hat \mu_2$ & $\hat \sigma$ & $\hat \xi$ & CV GLD   \\  \hline   
			Model 0   (Sta)  	  & 1.48 (.017) & & & .139 (.013) & .196 (.084) &  16.70  \\
			Model 1   (t)         & 1.39 (.037) & .0019 (.0006) &  & .125 (.010) & .120 (.085) &  15.95   \\
			Model 1 (t)$^*$       &1.38 (.03)& .0020 (.0005) & & .124 (.010) & .125 (.070) & \\
			Model 2   (SOI)       & 1.49 (.018) & .060 (.022) &  & .137 (.012) & .246 (.081) &  16.43   \\
			Model 3   (t+SOI)     & 1.34 (.033)& .0020 (.0006) & .064 (.021) & .122 (.010) & .169 (.075) & 15.54     \\
			\hline
		\end{tabular}
	\end{center}
	\vspace{.1 cm}
{\small *: Results using the maximum likelihood estimation provided by Coles (2001). Model 0: Stationary generalized extreme value model with L-moment estimates.}
\end{table}
 
\section{Discussion}\label{sec:diss}

In the simulation study, we observed that the MLE generally performed well, except for a few unusual samples, as depicted in the sample plots in the Supplementary Information.  In these cases, MLE is heavily influenced by outliers or large values near the end of the sample, resulting in inferior performance, especially for $\xi \le -.2$. Therefore, a robust MLE method (e.g., Wilcox 2021; Grego and Yates 2024) or a penalized MLE approach (Cannon 2010; Papukdee et al.~2022) could mitigate this problem. We defer this investigation to future research endeavors.

In solving a system of three equations in Step~2 of the proposed method, we encountered a very few case where the proposed method produced nonunique solutions in the root-finding algorithm. To address this problem, we selected one solution with a minimum value of $\chi$:
\begin{equation} \label{chi}
\chi = \sum_{i=T_1}^{T_{max}} \;\frac{ | E_n(i) -S_n(i) | }{  E_n(i)},
\end{equation}
where $T=(5, 10, 20, 40, T_{max})$, $E_n(i) = n/T_i$, and $S_n(i) = \sum_{j=1}^n I (x_j \ge q(T_i))$. Here, $I$ is the indicator function where $I(A)=1$ or $I(A)=0$ if $A$ is satisfied or not satisfied, respectively. 
Moreover, $q(T_i)$ is the RL corresponding to the return period $T_i$, calculated under the assumed model. In this study, we set $T_{max} = 80$ when $n=50$. Further, $E_n(i)$ and $S_n(i)$ represent the expected number of observations that are greater than the RL corresponding to $T_i$ without assuming any model and with the assumed model, respectively. Thus, the value in (\ref{chi}) serves as a distance measure for the expected number of events between the observations and the assumed model. 
%This measure ($\chi$) is considered by mimicking the well-known chi-square statistic, which measures the relative discrepancy between theoretical and observed numbers in the categorical data analysis. Instead of using the squared difference in the numerator, $\chi$ in (\ref{chi}) uses the absolute difference.

As with the transformation (\ref{transform}) to a standard Gumbel distribution in Step 2 of the proposed method, a similar transformation to the standard exponential distribution is available for the NS generalized Pareto model. Therefore, the proposed method for any NS model may be applicable whenever such a transformation to stationary sequences is available. However, Gumbel’s choice is arguably the most natural, given its status within the GEV family {(Coles 2001, pp.~111).} Nonetheless, one can choose a unit Fréchet transformation: $ Z^*_t = {{-1} / \ {\text{log} \; \hat F_t (Z_t)} }$, where $\hat F_t$ is the estimated CDF of the NS GEV model. %Initially, we thought this transformation appeared more appropriate than Gumbel’s choice because unit Fréchet has three parameters. Thus, $Z^*_t$ would be more useful for the three-parameter estimation in Step 2 of the proposed method. 
However, $Z^*_t$ still depends on the estimated shape parameter $\hat \xi$; thus, we hardly obtain the solution of a system of equations (\ref{LME-NS}) without fixing $\hat \xi$. %Fixing $\hat \xi$ may not improve the estimation performance. 
Therefore, we conclude that the unit Fréchet transformation is unsuitable for the proposed algorithm. 

%We tried to figure out the reason why there are multiple solutions in a system of equations in the Step 3 of the proposed method, even though the multiple solutions happen in a few cases. It may be because the proposed method attempts to find three estimates by using two parameters Gumbel distribution. We need further study for those cases.
	
Motivated by Jan et al.~(2021), we applied trimmed L-moments (Elamir and Scheult 2003) to the proposed method. However, the results of the limited simulation experiment using the trimmed L-moments did not exhibit significant improvement over the proposed method without trimming. Nevertheless, a trimmed L-moment approach can be helpful for samples with outliers.

%Although it is inevitable to fit an NS GEV model to the data and estimate future RLs based on the fitted NS model, caution must be exercised when projecting trends in a time-dependent NS GEV model. The trend observed in the model may not continue in the future (Obeysekera and Salas 2014), and the uncertainty associated with the estimated parametric function should be higher in the future than in the past. 

\section{Conclusion}
Extreme values are inherently rare events, necessitating an effective method for fitting models using a few extreme observations. The L-moment method is more efficient than the MLE method for small samples. Therefore, this study proposed a method combining robust regression and L-moments to fit NS GEV models to time-series data. 
The proposed method improved the NS GEV model by addressing a weakness of the GN16 method, particularly regarding the GEV11 model with increasing variance. %The GEV11 model is an NS GEV model with a time-dependent linear function in the location parameter and a time-dependent log-linear function in the scale parameter, as represented by (\ref{mu_t}) and (\ref{sigma_t}), respectively. 
%The simulation study, especially for the GEV11 model, revealed that the proposed method outperforms the MLE, stationary LME, and WLS methods. 
To illustrate the usefulness of the proposed method, we provided a real application using the extreme streamflow data from Trehafod in the UK. We extended the proposed method to an NS GEV model with physical covariates. Furthermore, we considered a model selection criterion based on the cross-validated generalized L-moment distance as an alternative to the likelihood-based criteria.

The likelihood-based method is often straightforward to apply in complex situations. In contrast, the L-moment-based method is not as straightforward. Thus, establishing an L-moment-based procedure for complex situations is challenging because the L-moment method is more efficient than the MLE method for small samples (of extreme values). This study is meaningful in that we considered developing a new methodology based on L-moments to build improved NS extreme value models.

\section*{Appendix}
\subsection*{Estimation of each parameter in the WLS method}

For further analysis from the WLS method, such as for calculating the standard error or for testing hypothesis on each parameter, %or in estimating the redefined RL by Parey et al.~(2010), 
we must obtain the final estimates for each parameter.
In GEV11 model, we denote the final estimates as $\hat\mu_0^f,\; \hat\mu_1^f,\; \hat\sigma_0^f,\; \hat\sigma_1^f,\; \hat\xi^f$. To obtain these final estimates, we set the right-hand side of (\ref{SKq}) to be the same as the effective RL (\ref{conv_rt}). Then, by solving this equation with respect to the five parameters, we have the following: 
\begin{equation}\label{SK_final}
\hat\sigma_0^f = \hat\sigma_0 +\text{log} \;\hat \sigma_{st}, ~~~ ~
\hat\sigma_1^f = \hat\sigma_1, ~~~~ \hat\xi^f = \hat\xi_{st}.
\end{equation}
However, $\hat\mu_0^f$ and $\hat \mu_1^f$ are not directly obtained because 
\begin{equation} \label{mu1_f}
\hat\mu_0^f + \hat\mu_1^f \times t = \hat\mu_0 + \hat\mu_1 \times t + \hat\mu_{st} \times \text{exp}( \hat\sigma_0 + \hat\sigma_1 \times t ). 
\end{equation}
To obtain approximated values of $\hat\mu_0^f$ and $\hat \mu_1^f$ in this study, we fitted the right-hand side of (\ref{mu1_f}) to the simple regression for the time variable over the range $t=1,2,\cdots,n$. Then, the estimated regression coefficients are approximated values of $\hat\mu_0^f$ and $\hat \mu_1^f$. 

However, for some purposes such as computing T-year return level in (\ref{conv_rt}) or (\ref{Parey_rt}), one can use the right-hand side of (\ref{mu1_f}) directly as $\mu_t$, instead of this approximation. {Denote $\hat\mu_t^f, \; \hat\sigma_t^f,\; \hat\xi^f$ be the final estimates of NS parameters of GEV($\mu_t, \, \sigma_t,\, \xi_t$). Then 
	\begin{eqnarray} \label{final_para}
	\hat \mu_t^f &=& \hat\mu_0 + \hat\mu_1 \times t + \hat\mu_{st} \times \text{exp}( \hat\sigma_0 + \hat\sigma_1 \times t ) , \\
	\hat \sigma_t^f &=& \exp\, (\, \hat\sigma_0 +\text{log} \;\hat \sigma_{st} + \hat\sigma_1 \times t), ~~~~
	\hat \xi^f  = \hat\xi_{st}.
	\end{eqnarray}
Then we apply these estimates directly to (\ref{conv_rt}) or (\ref{Parey_rt}).
}

For the GEV10 and GEV20 models, to obtain the final WLS estimates, we set the right-hand side of (\ref{SKq}) to be the same as the effective RL (\ref{conv_rt}). Then, by solving this equation with respect to the parameters, we obtain the final WLS estimates for the GEV10 model:
\begin{equation}\label{SK10_final} 
\hat\mu_0^f = \hat\mu_0 + \hat \mu_{st} \times \hat\sigma, ~~~ ~
\hat \mu_1^f = \hat \mu_1,  ~~~~
\hat\sigma^f = \hat\sigma \times \hat \sigma_{st}, ~~ ~~
\hat\xi^f = \hat\xi_{st}.
\end{equation}
For the GEV20 model, we obtain the final WLS estimate $\hat \mu_2^f = \hat \mu_2$, while other parameters are the same as in (\ref{SK10_final}).

\begin{small}
	\renewcommand{\baselinestretch}{0.8}
	\subsection*{Funding and acknowledgments}
	
	 The authors thank the reviewers and the associate editor for their valuable comments and constructive suggestions that considerably improved the manuscript.  
	 This work was supported by %BK21 FOUR (No.5120200913674) and 
	 the NRF grant (RS-2023-00248434) funded by 
	 %the Ministry of Education and 
	 the National Research Foundation of Korea and by Global - Learning \& Academic research institution for Master's, PhD students, and Postdocs(LAMP) Program of National Research Foundation of Korea(NRF) grant funded by Ministry of Education (No.~RS-2024-00442775).
	
	\subsection*{Code and data availability}
	
	Peak streamflow data in Trehafod, UK: https://nrfa.ceh.ac.uk/peak-flow-dataset.\\
	R code: https://github.com/yire-shin/non\_stationary-gev.git
	
	\subsection*{Conflict of interest}
	The authors declare no potential conflicts of interest.
	
	\subsection*{ORCID}
	Yire Shin, 0000-0003-1297-5430; $~~$
	Yonggwan Shin, 0000-0001-6966-6511 \\
%	Sanghoo Yoon, 0000-0003-4790-0148; $~~$
	Jeong-Soo Park, 0000-0002-8460-4869 
	
	\section*{Author contributions statement}
	Yire Shin and Jeong-Soo Park conceived the study, Yonggwan Shin conducted the analysis. All authors wrote the first draft, reviewed, edited, and approved the final manuscript.
	
\end{small}

%%%%%%%%%%%%%%%%%%%% References %%%%%%%%%%%%%%%%%%%%%%%
%\bibliography{References1-11.bib}

\renewcommand{\baselinestretch}{0.8}

\end{document}